\documentclass[final,3p,times,twocolumn]{elsarticle}
 \biboptions{comma,sort&compress}
\usepackage{here}
 \usepackage{graphicx}
  \usepackage{epsfig}

\def\nin{\noindent}
\def\beq{\begin{equation}}
\def\eeq{\end{equation}}
\def\bea{\begin{eqnarray}}
\def\eea{\end{eqnarray}}

\journal{Nuc. Phys. (Proc. Suppl.)}

\begin{document}

\begin{frontmatter}



\title{Top Quark Measurements in CMS}

 \author[label1]{Efe Yazgan\corref{cor1}}
  \address[label1]{Department of Physics and Astronomy, Ghent University, 
\\
Proeftuinstraat 86 B-9000 Gent, Belgium.}
\cortext[cor1]{On behalf of the CMS Collaboration}
\ead{efe.yazgan@cern.ch}



\begin{abstract}
\noindent
Measurements involving top quarks provide important tests of QCD. A selected set of top quark measurements in CMS including the strong coupling constant, top quark pole mass, constraints on parton distribution functions, top quark pair differential cross sections, $t\overline{t}$+0 and $>0$ jet events, top quark mass studied using various kinematic variables in different phase-space regions, and alternative top quark mass measurements is presented. The evolution of expected uncertainties in future LHC runs for the standard and alternative top quark mass measurements is also presented. 

\end{abstract}


\end{frontmatter}


\section{Introduction}
\nin
The top quark is the most massive particle known to date. 
The dominant mechanism of top quark production at hadron colliders is $t\overline{t}$ pair production through QCD interactions. At the LHC, gluon fusion is the dominant mechanism for $t\overline{t}$ pair production . Single top quark production happens through electroweak interactions. 
The top quark decays via the weak interaction almost exclusively to a b quark - W boson pair before hadronisation that occurs at a typical time-scale of $1/\Lambda_{QCD}$. Therefore, the properties of the top quark, before being obscured by QCD effects, can be measured. 
The top quark mass combined with the W boson and Higgs boson masses completes the Standard Model (SM). Moreover, the top quark and the Higgs boson mass measurements can be used to obtain hints about the stability of the vacuum \cite{shaposhnikov2012, Degrassi2012, Moch2014}. 
Top quark measurements test perturbative QCD. Moreover, understanding perturbative and non-perturbative QCD effects is crucial to obtain the ultimate precision in top quark mass and its interpretation. In this proceeding a selection of measurements from CMS \cite{CMS} are summarised that are most relevant for QCD. 
\section{Determination of the strong coupling constant and the top quark pole mass from $t\overline{t}$ cross section}
\begin{figure}[hbt] 
\centerline{\includegraphics[width=7.5cm]{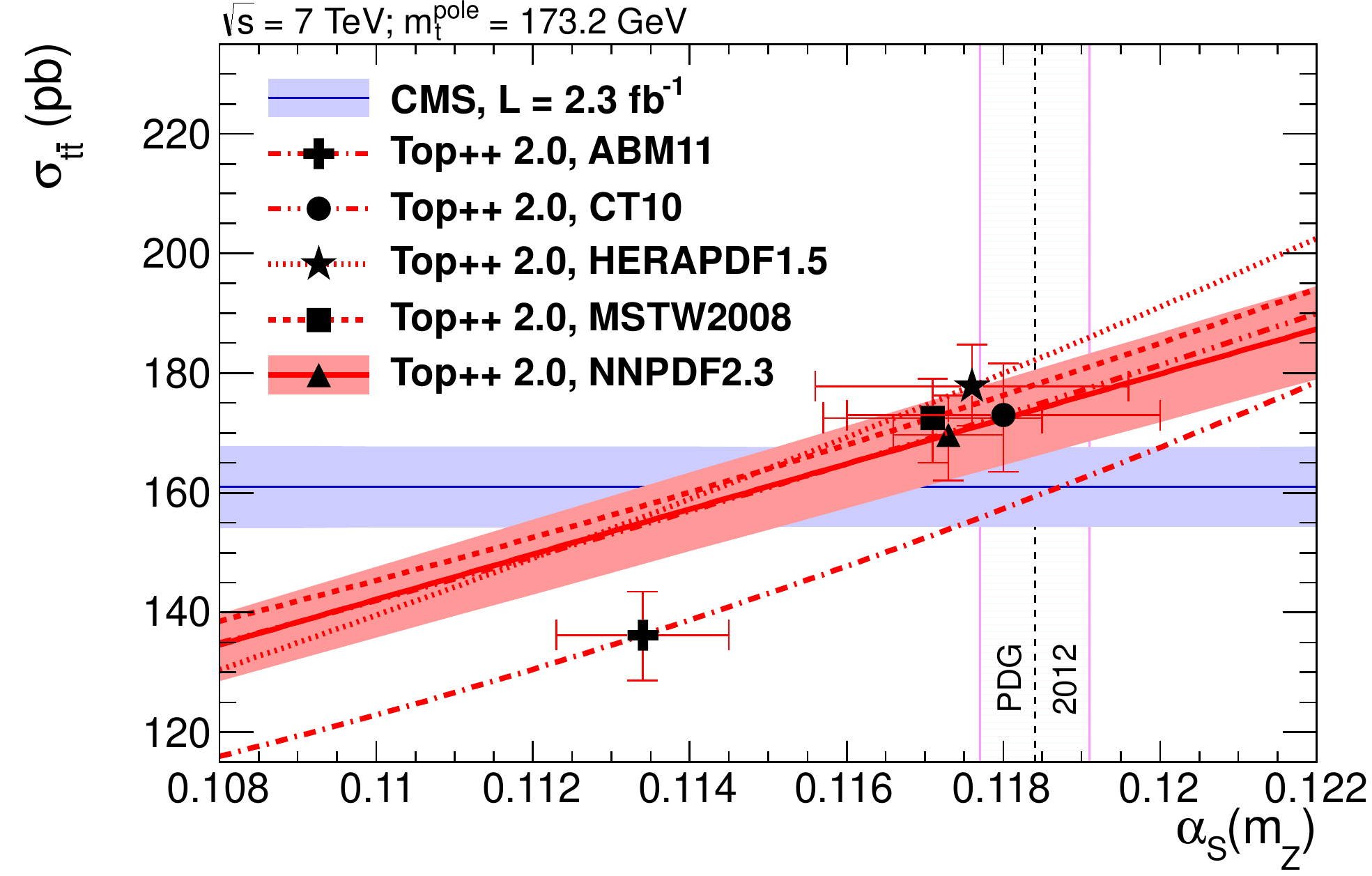}}
\caption{\scriptsize $t\overline{t}$ cross section vs the strong coupling constant. The theoretical cross section predictions at NNLO+NNLL are given for five different NNLO PDF sets.The uncertainties on the the
measured  cross section is shown with a blue band around the parametrised CMS cross section measurement. The renormalisation and factorisation scale and PDF uncertainties on the theoretical prediction using the NNPDF2.3 PDF set is shown as a blue band. The latest world average for the strong coupling constant is displayed as a hatched band.}
\label{foldedLikelihoods_pole_xsec_vs_alpha} 	
\end{figure}

\begin{figure}[hbt] 
\centerline{\includegraphics[width=7.5cm]{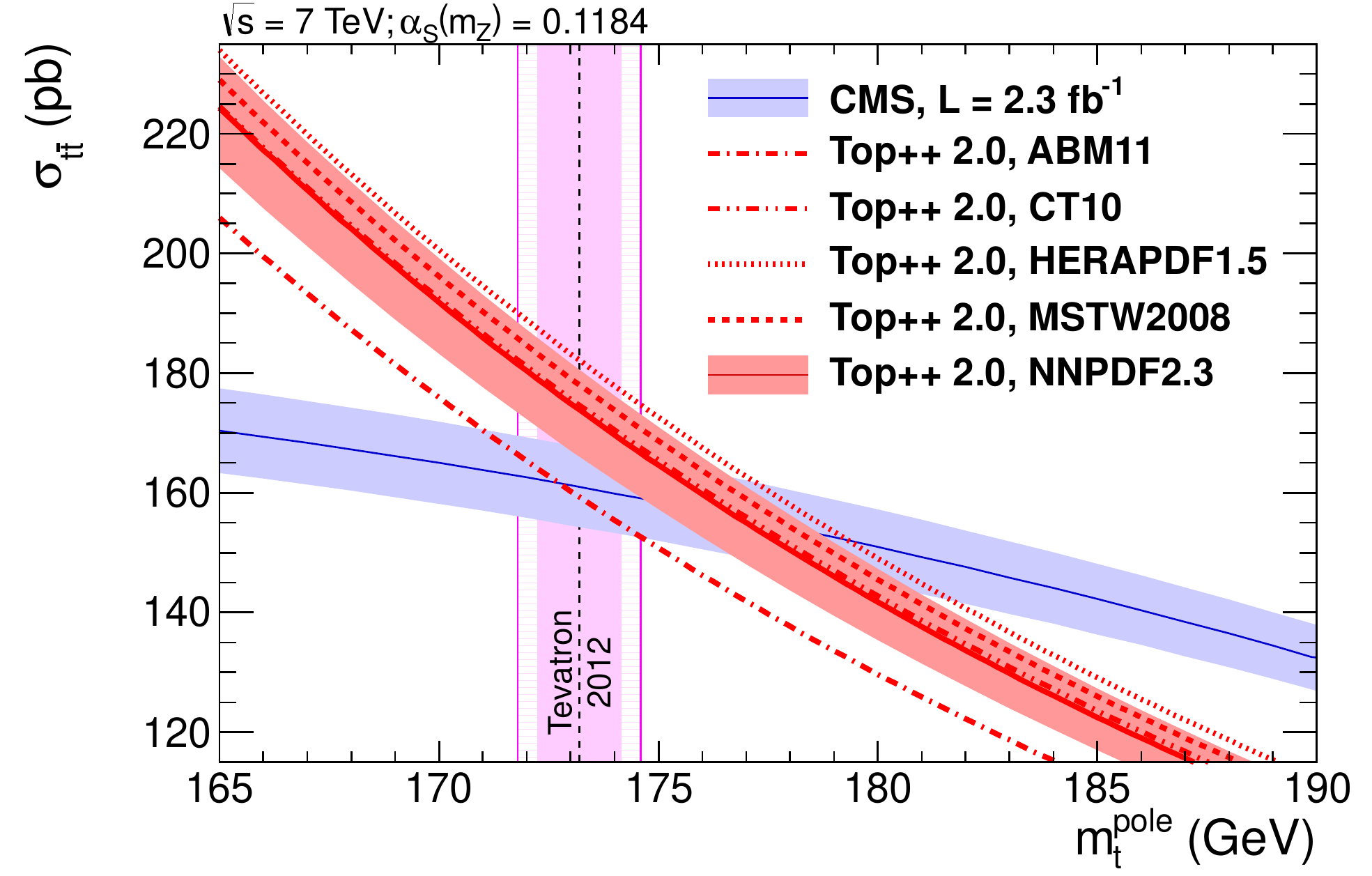}}
\caption{\scriptsize $t\overline{t}$ cross section vs the top quark pole mass. The theoretical cross section predictions at NNLO+NNLL are given for five different NNLO PDF sets.The uncertainties on the the
measured  cross section is shown with a blue band around the parametrised CMS cross section measurement. The renormalisation and factorisation scale and PDF uncertainties on the theoretical prediction using the NNPDF2.3 PDF set is shown as a blue band. The Tevatron average for the top quark mass is displayed as a hatched band. The outer band indicates an additional uncertainty to account for the possible difference between the direct mass measurements from Tevatron and the top quark pole mass.}
\label{foldedLikelihoods_pole_xsec_vs_mass} 
\end{figure}

\nin
The free parameters of the QCD Lagrangian are the quark mass values and the strong coupling constant, $\alpha_s$. 
Along with the Parton Distribution Functions (PDFs), these constitute the external inputs in the $t\overline{t}$ cross-section calculation.
The strong coupling constant, $\alpha_s$,  depends on the energy scale of the hard process and its evolution is determined by the renormalisation group equations. Many different processes are used to determine $\alpha_s$  among which lattice calculations and $\tau$-decays yield the most precise results \cite{pdgalphas2012}. At the LHC, using jet data, $\alpha_s$ is measured up to a scale of 1.4 TeV \cite{alphasjet}.  However, this is determined only up to NLO QCD (or approximate NNLO) with large scale uncertainties and possible unaccounted non-perturbative corrections. Comparing the most precise single $t\overline{t}$ cross-section measurement at  $\sqrt{s}=7$ TeV \cite{dilepton7TeVttbarcs} with a NNLO+NNLL calculation, CMS made a determination of $\alpha_s$ constraining $m_t$ and also of $m_t$ constraining $\alpha_s$ \cite{alphasmtpole}. The $t\overline{t}$ cross section predicted from a calculation at NNLO+NNLL using different NNLO PDF sets are shown as a function of $\alpha_s(m_Z)$ in Figure \ref{foldedLikelihoods_pole_xsec_vs_alpha} and $m_t^{pole}$ in Figure \ref{foldedLikelihoods_pole_xsec_vs_mass}. The dependence of the cross-section on $m_t^{pole}$ and $\alpha_s(m_Z)$ originates from the change in the event kinematics which in turn modifies the needed acceptance corrections to obtain the cross section.  
Fully inclusive calculations at NNLO+NNLL QCD are given for five different NNLO PDF sets as a function of these variables. Using the NNPDF2.3 set, $\alpha_s(m_Z)$ is measured to be 0.1151$^{+0.0028}_{-0.0027}$ by fixing the $m_t^{pole}$ value to the Tevatron average of 173.18$\pm$0.94 GeV \cite{tevmass2013}. The dominant systematic uncertainty sources are factorisation and renormalisation scales, top quark mass value, and LHC beam energy. This is the first $\alpha_s(m_Z)$ measurement using $t\overline{t}$ events and the first determination of this quantity at full NNLO QCD at a hadron collider. Similarly, fixing  $\alpha_s(m_Z)$ to 0.1184$\pm$0.0007 \cite{pdgalphas2012}, a top quark pole mass of 176.7$^{+3.0}_{-2.8}$ GeV is measured.  The dominant systematic uncertainties are due to the uncertainty on the measured $t\overline{t}$ cross-section and PDF uncertainties. This mass measurement method tests the mass scheme used in Monte Carlo simulations. Moreover, it provides complementary and different systematic uncertainties than the ones in the direct top quark mass  measurements.  A similar measurement of the top quark mass is made by the ATLAS collaboration \cite{atlaspole2014} and comparisons have been made to D0 \cite{D0pole2009} and CMS top quark pole mass measurements as well as to the world average. The top quark mass values extracted in different experiments using different methods and assumptions are found to be consistent.

\nin
\section{Constraints on PDFs from top quark pair production}
\nin
Czakon et al. \cite{czakon2013} have shown that including the LHC top quark data in the PDF fits, the uncertainty on the gluon PDFs can be reduced up to $\sim25\%$ depending on the parton momentum fraction, x. This will have significant impact on the predictions for the Higgs boson and many beyond standard model processes that are dominated by the gluon-fusion process at the LHC. 

\begin{figure}[hbt] 
\centerline{\includegraphics[width=7.5cm]{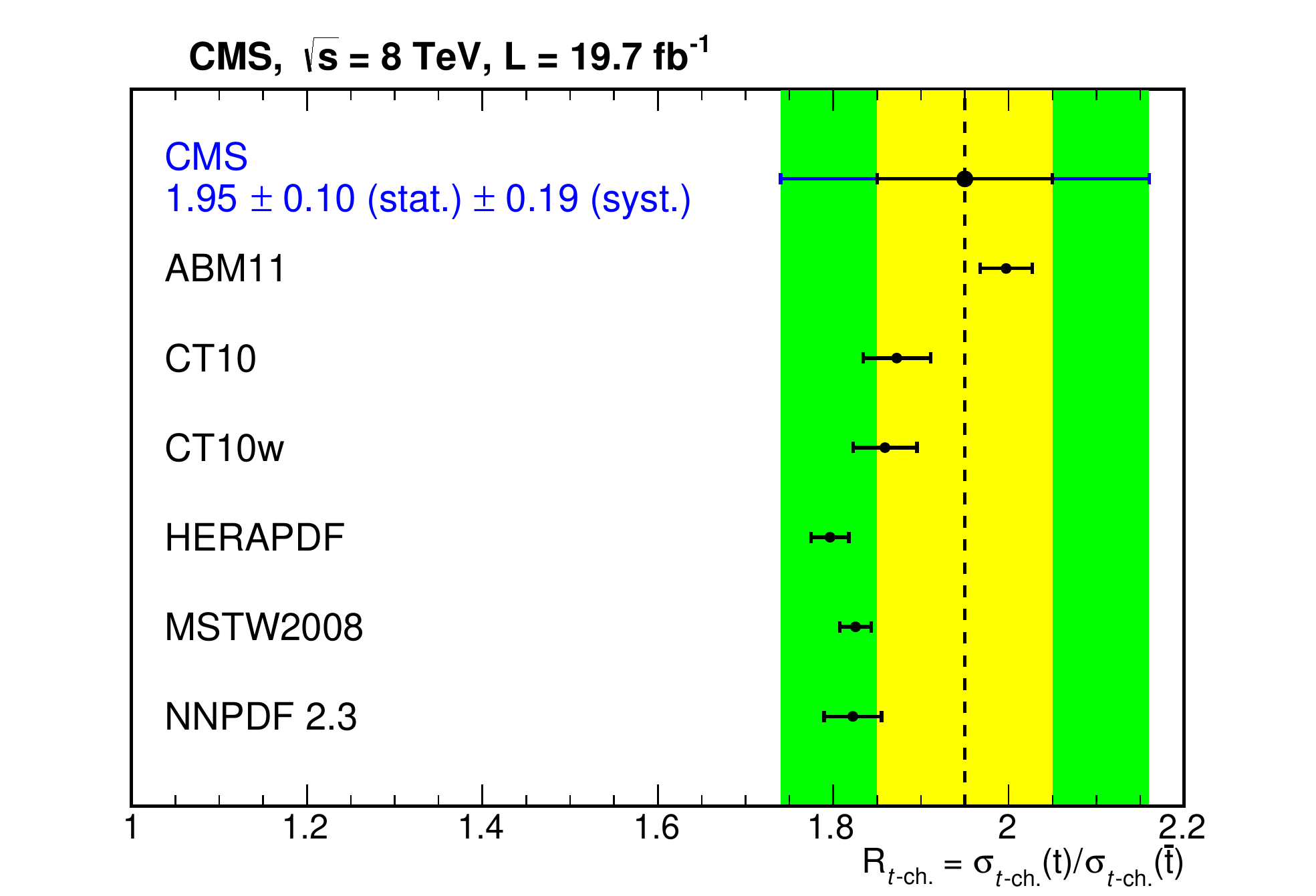}}
\caption{\scriptsize Measured $R_{t-ch.}$ and predictions from several PDF sets. The statistical error bar on data is shown with yellow and systematic uncertainty is shown with green. The error bars on the predictions from PDF sets comprise the uncertainties due to factorisation and normalisation scales, top quark mass and the statistical uncertainty.}
\label{fig3} 
\end{figure}

In single top quark production in the t-channel, top and anti-top quark cross sections are different because of the up- and down-type quarks in their initial states respectively. 
This makes the ratio of the top and anti-top quark cross-sections ($R_{t-ch.}=\sigma_t/\sigma_{\overline{t}}$) sensitive to the $u$ and $d$  PDFs of the proton. The ratio, $R_{t-ch.}$, probes the couplings in $Wtb$ \cite{Aguilar-Saavedra2008}, as well as flavour-changing neutral currents \cite{gao2011}. 
$R_{t-ch.}$ is predicted to be $\sim2$ in the proton PDFs.  CMS measured the  t-channel, top and anti-top quark cross sections, as well as their ratio, at a center of mass energy of $\sqrt{s}=8$ TeV. The measured ratio is $R_{t-ch.}$=1.95$\pm$(stat.)$\pm$0.19(sys.)  \cite{singletoptudpdf}. The measurement along with predictions from different PDF sets is shown in Figure \ref{fig3}. 
All measurements are found to be consistent with the SM calculations. However, it is observed that not all PDF sets are compatible with each other. A precise measurement of 
$R_{t-ch.}$ may discriminate between different PDF sets. Moreover, the ratio of 8 and 7 TeV cross-section measurements will potentially provide complementary information on the PDFs.  
\section{Top quark pair differential cross sections}
\begin{figure}[hbt] 
\centerline{\includegraphics[width=7.5cm]{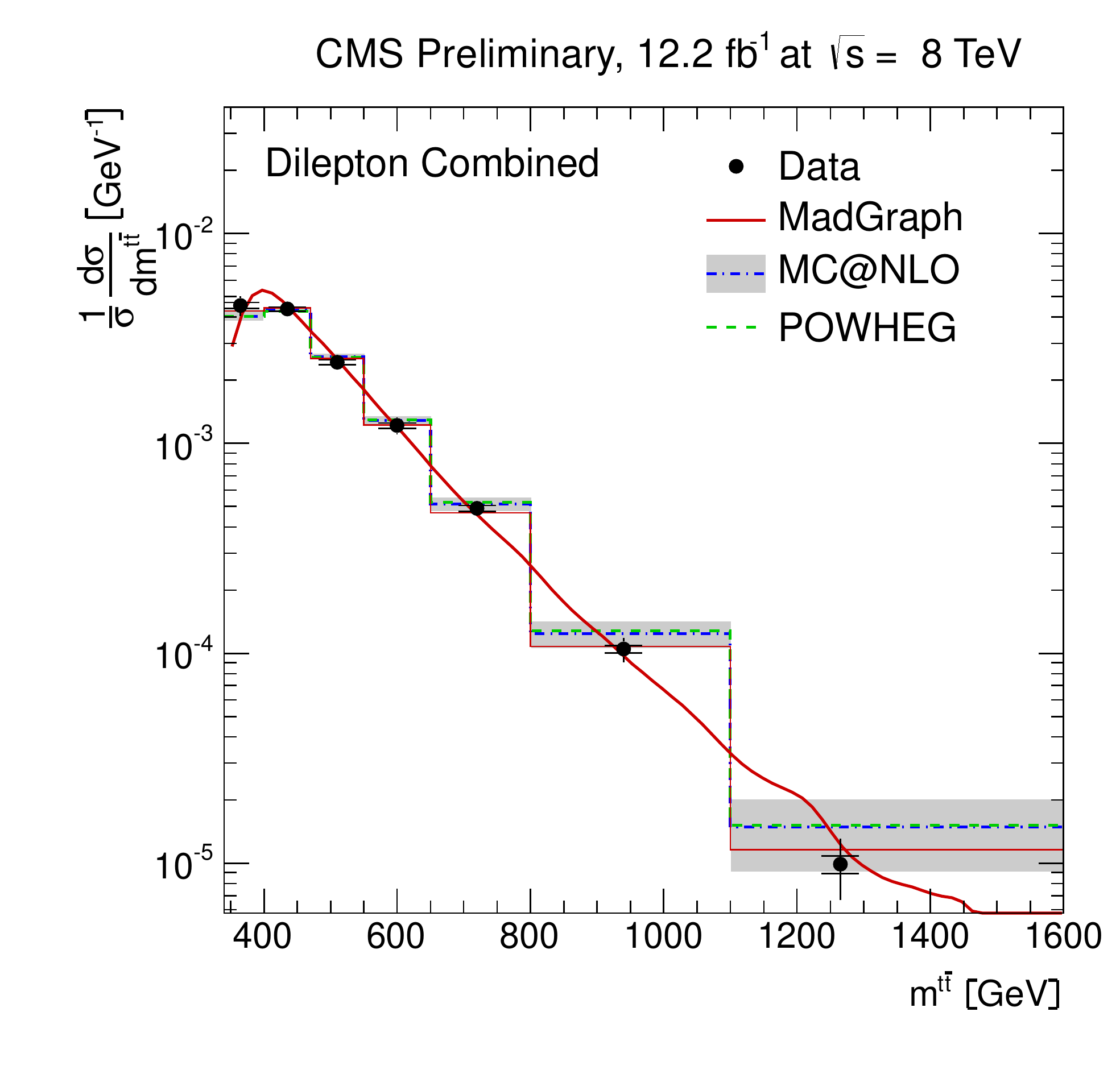}}
\caption{\scriptsize Normalized differential ttbar production cross section as a function of the invariant mass of the top-antitop quark pair. The invariant mass is presented at the parton level extrapolated to the full phase space. The inner (outer) error bars indicate the statistical (combined statistical and systematic) uncertainty. The data is compared to the estimations from MadGraph, POWHEG, and MC@NLO Monte Carlo generators. The prediction from MadGraph is displayed as a curve along with a binned histogram.}
\label{figdiff1} 
\end{figure}

\begin{figure}[hbt] 
\centerline{\includegraphics[width=7.5cm]{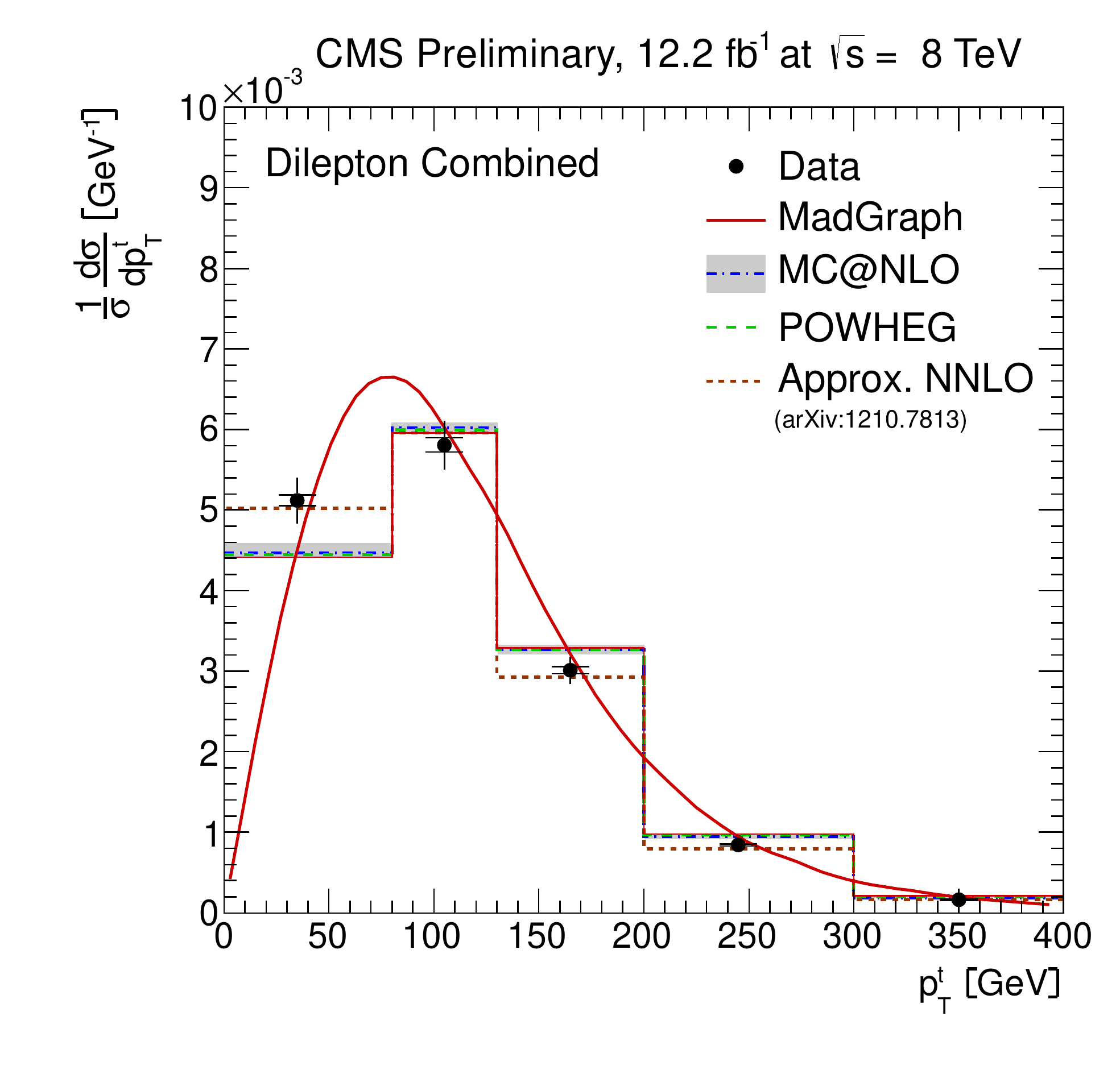}}
\caption{\scriptsize Normalized differential ttbar production cross section as a function of the transverse momentum of the top or the anti-top quark. The transverse momentum is presented at the parton level extrapolated to the full phase space. The inner (outer) error bars indicate the statistical (combined statistical and systematic) uncertainty. The data is compared to the estimations from MadGraph, POWHEG, and MC@NLO Monte Carlo generators as well as to an approximate NNLO calculation. The prediction from MadGraph is displayed as a curve along with a binned histogram.}
\label{figdiff2} 
\end{figure}

\nin
Measurements of top quark pair differential cross sections test the validity of various levels of perturbative QCD approximations for top quark production, test and tune MC models. The tails of the differential distributions can be used in new physics searches. In the CMS measurements \cite{topdiffljets,topdiffdilepton} distributions of leptons and b-jets are corrected to parton level within kinematic and geometric acceptance as well as to the full phase-space to be able to compare to approximate NNLO predictions. Two of the corrected distributions are displayed in Figures \ref{figdiff1} and \ref{figdiff2}. All measurements in different channels agree with each other and with SM predictions.
A discrepancy of the predictions from different MC generators with data is observed in the transverse momentum distribution of the top quarks (see Figure \ref{figdiff2}) although the data agrees well with the approximate NNLO predictions. Such discrepancies with MC generators need to be understood and are taken into account in measurements and search analyses. 
\section{Measurement of jet multiplicity distributions in $t\overline{t}$ production}
\begin{figure}[hbt] 
\centerline{\includegraphics[width=7.5cm]{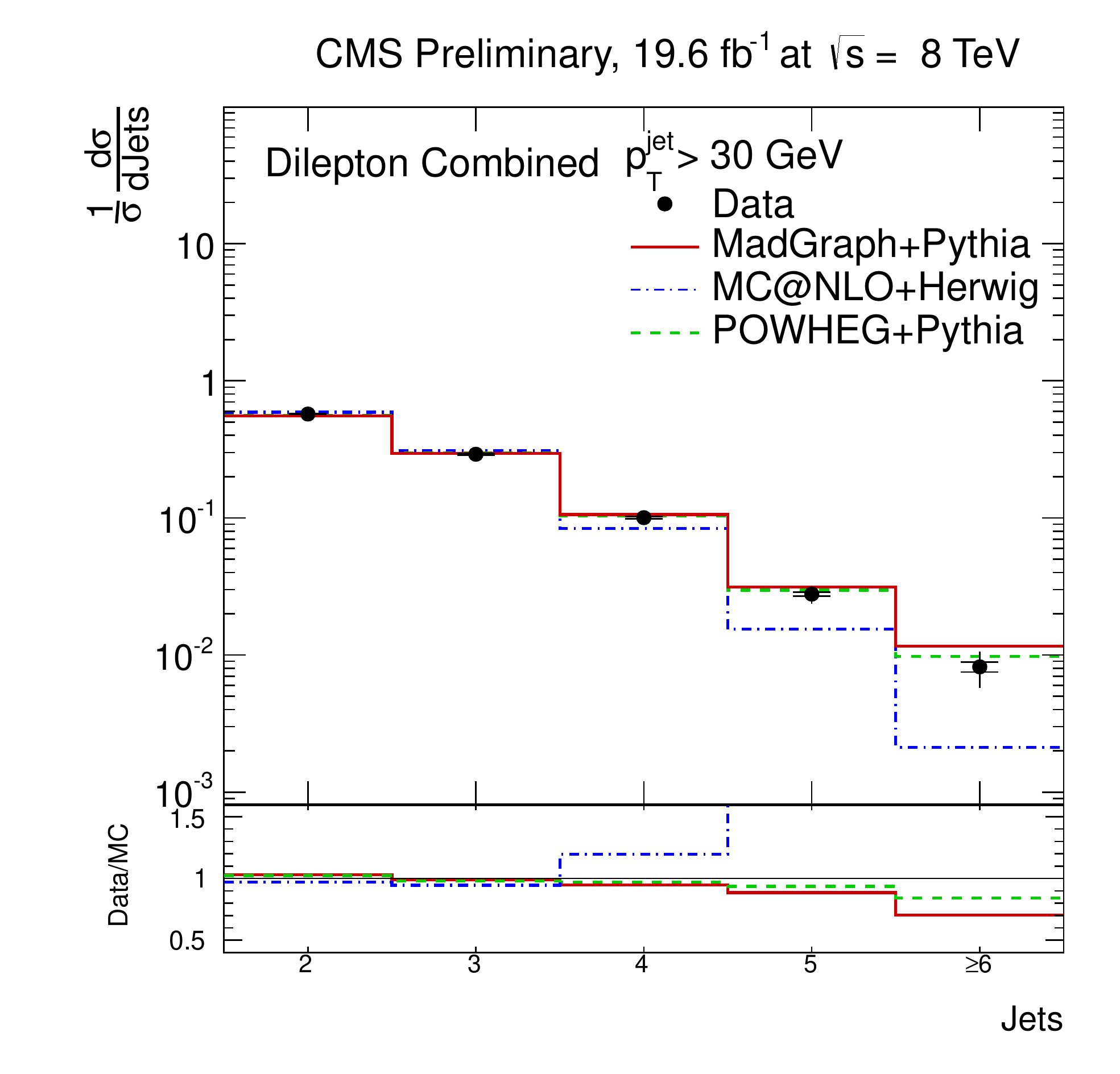}}
\caption{\scriptsize  Normalised differential cross-section vs. jet multiplicity for jets with $p_T > $ 30 GeV and predictions from different MC generators. The inner error bar represents the statistical error and the outer error bar represents the total error.}
\label{fig4.51} 
\end{figure}

\begin{figure}[hbt] 
\centerline{\includegraphics[width=7.5cm]{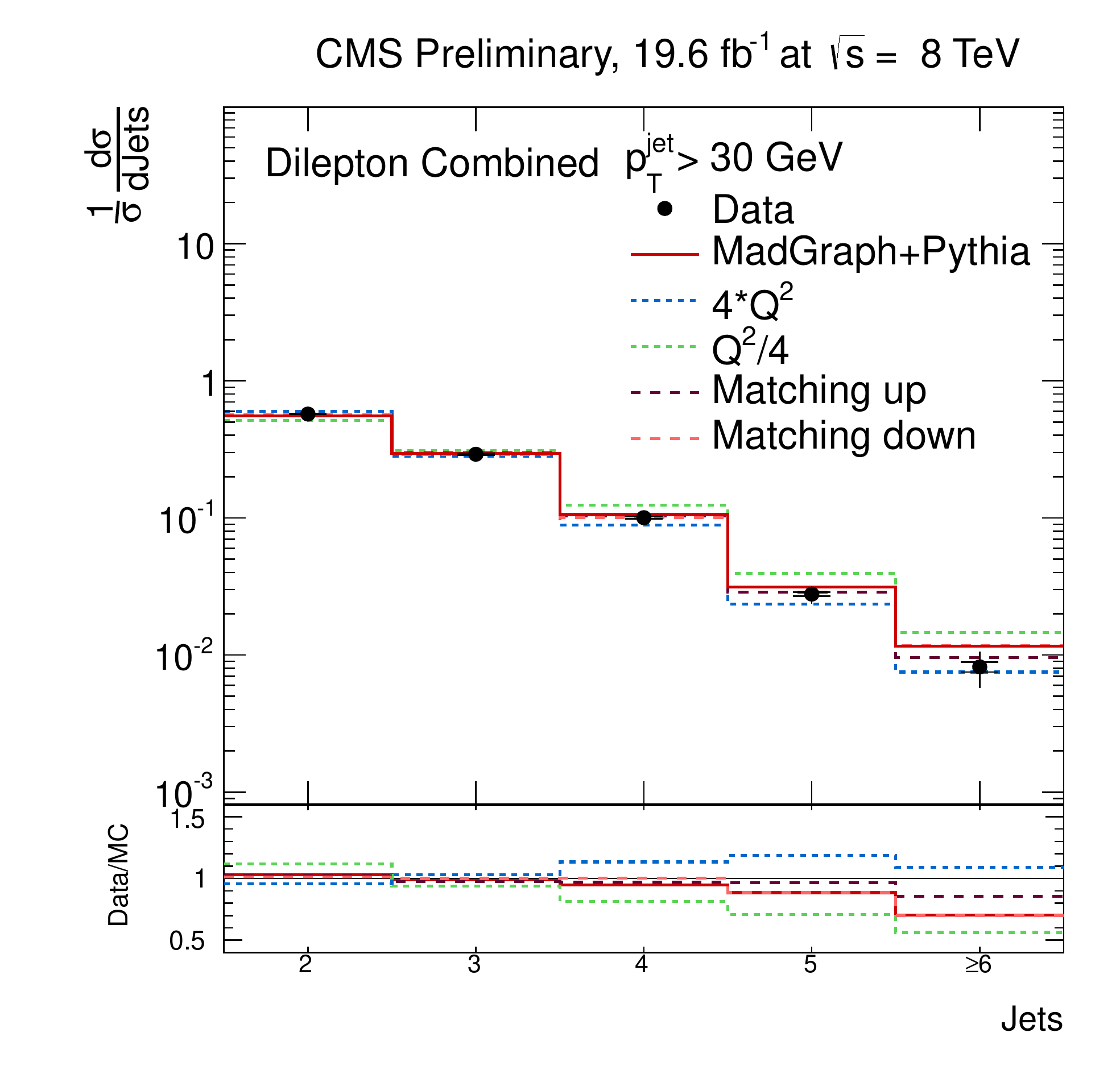}}
\caption{\scriptsize  Normalised differential cross-section vs. jet multiplicity for jets with $p_T > $ 30 GeV compared to the predictions from MadGraph with varied factorisation x normalisation scale and jet-parton matching threshold. The inner error bar represents the statistical error and the outer error bar represents the total error.}
\label{fig4.52} 
\end{figure}

\nin
At the LHC, $\sim50\%$ of $t\overline{t}$ pairs are accompanied by additional hard jets from initial or final state QCD radiation (ISR/FSR). Measurements of these additional jets in $t\overline{t}$ production test higher-order QCD calculations, and modelling of ISR/FSR. Moreover, $t\overline{t}$+jets events is an important background to Higgs boson measurements particularly in the $H\rightarrow b\overline{b}$ channel. An abnormal number of jets distributions may be an indication of new physics. Calculations of $t\overline{t}+\leq2$ jets are available at NLO QCD.   
CMS made measurements of the jet  multiplicity distributions in $t\overline{t}$ production at $\sqrt{s}$=7 and 8 TeV \cite{ttjets1,ttjets2}. 
Figure \ref{fig4.51} displays the normalised differential cross-section vs. jet multiplicity for jets with $p_T > $ 30 GeV compared to predictions from MadGraph+Pythia, MC@NLO+Herwig and Powheg+Pythia. It is observed that 
MC@NLO+Herwig does not describe the jet multiplicity distribution for events with more than four jets. 
Figure \ref{fig4.52} displays normalised differential cross-section vs. jet multiplicity for jets with $p_T > $ 30 GeV compared to the predictions from MadGraph with varied factorisation x normalisation scale and jet-parton matching threshold. It is observed that 
MadGraph with smaller factorisation x renormalisation scale yields a worse description of the data. 

\begin{figure}[h] 
\centerline{\includegraphics[width=7.5cm]{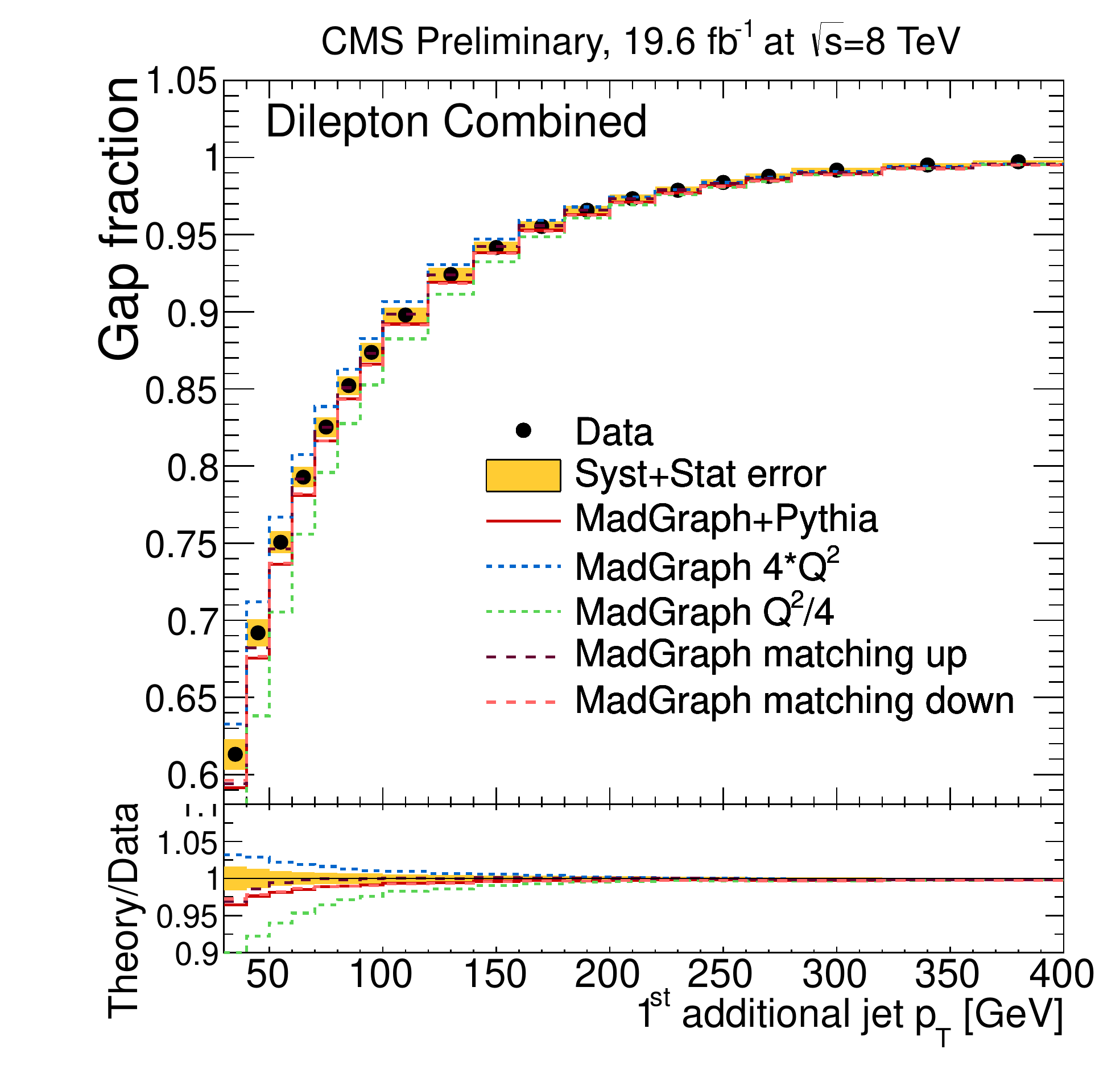}}
\caption{\scriptsize Gap fraction vs. the leading additional jet $p_T$. Data are compared to predictions from MadGraph+Pythia with various different $Q^2$ and jet-parton matching scales. The data points are shown with statistical error bars and the shaded band indicate the statistical and systematic errors added in quadrature.}
\label{fig5} 
\end{figure}

An alternative way to investigate the jet activity from QCD radiation is measuring the gap fraction defined to be the fraction of events that do not contain additional jets above a $p_T$ threshold. Figure \ref{fig5} shows the gap fraction vs. the leading additional jet $p_T$ compared to predictions from MadGraph+Pythia with difference scale choices. It is found that all tested MCs describe the gap fraction vs. second additional jet $p_T$ well. MC@NLO+Herwig describes the gap fraction vs the first additional jet $p_T$ in the event better and MadGraph with decreased $Q^2$ scale predicts lower gap fraction values than those of the observed ones. These are consistent with the measurements of $t\overline{t}$+jets measurements described above. 

\section{The top quark mass}
\begin{figure}[h!] 
\centerline{\includegraphics[width=7.5cm]{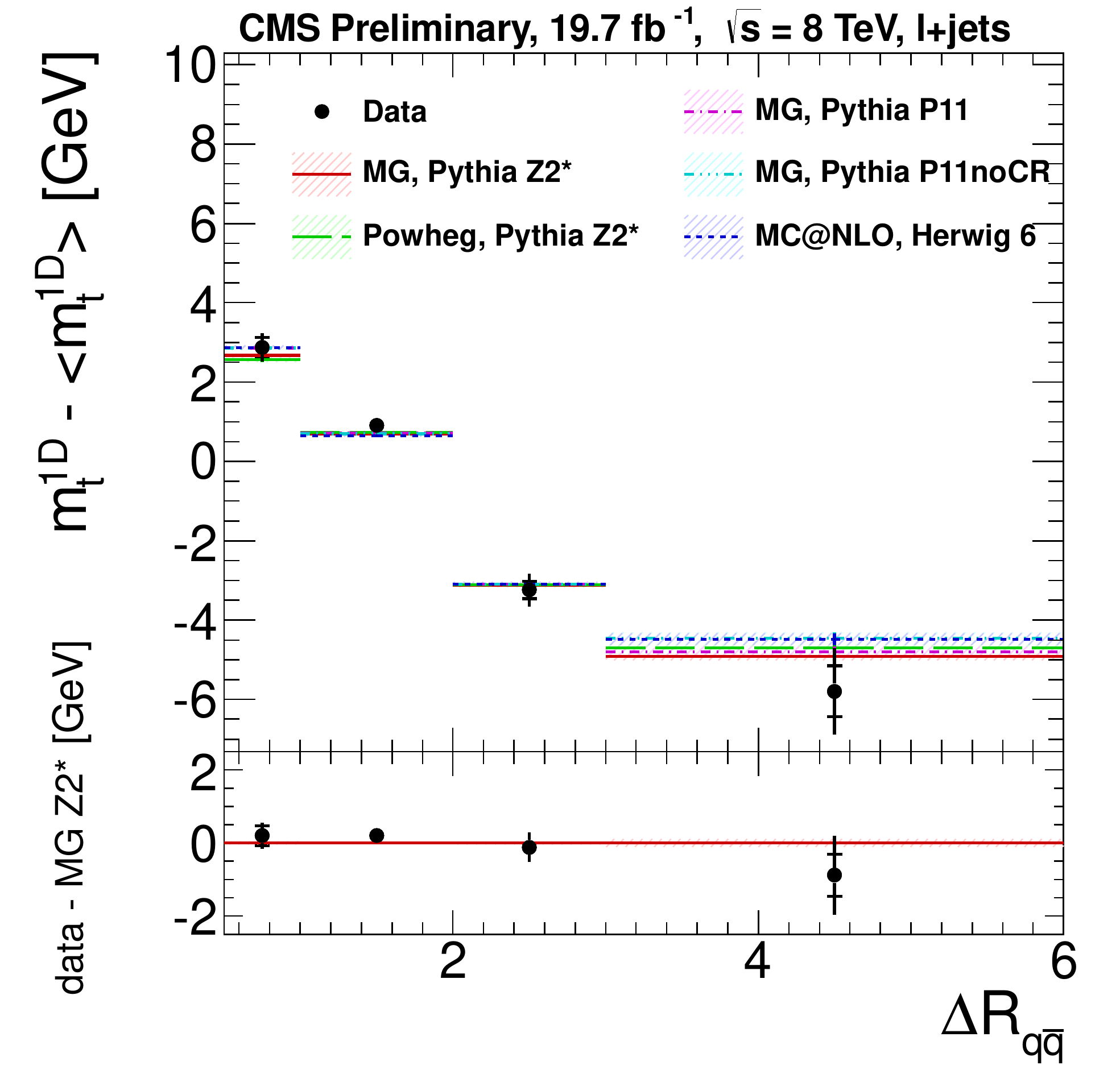}}
\caption{\scriptsize Dependence of top quark mass (measured in the 1D analysis) on $\Delta R_{q\overline{q}}$. The systematic and statistical errors are added in quadrature and the statistical error on the $t\overline{t}$ simulation is indicated as hatched areas. Beneath the plot is shown the difference between data and MC.}
\label{fig6} 
\end{figure}

\nin
Top quark mass ($m_t$) is a free parameter of the QCD Lagrangian, therefore it is not an observable and has no unique interpretation. 
In experiments, it has been measured in different ways. The top quark mass from direct measurements in colliders calibrated using MC simulations (hereon denoted by $m_t^{MC}$) is neither equivalent to the {\it pole mass} nor the {\it running mass} defined in a renormalisation scheme. This is because the measurements at colliders depend on MCs with matrix elements at fixed order (LO or NLO) QCD with higher orders being simulated by parton showers. In direct $m_t$ measurements usually each jet is assigned to a top quark decay product constrained by kinematic fits. The mass and jet energy scale (JES) are determined simultaneously in order to minimise the uncertainty from JES which is usually the dominant uncertainty source. The method is "calibrated" for biases using MC simulations. 
In direct $m_t$ measurements dominant uncertainties are flavour dependent jet energy scale, hadronisation and factorisation scale uncertainties. 
The world average with inputs from lepton+jets, di-lepton, all jets, and missing $E_T$+jets final states yield $m_t=173.34\pm0.27~(stat)\pm0.24~(JES)\pm0.67~(sys)$ GeV \cite{mtworld}. This is the first ever Tevatron-LHC combination. 
Using the ideogram method in the lepton+jets final state, CMS at $\sqrt{s}=8$ TeV obtained $m_t=172.04\pm0.19~(stat+JSF)\pm0.75~(sys)$ GeV and $JSF = 1.007\pm0.002~(stat)\pm0.012~(sys)$ \cite{mtideo8} which is the first single measurement with a precision less than 1 GeV. 
This precision is only a few times $\Lambda_{QCD}$ and less than the top quark width where the measurements and interpretation of the top quark mass become more challenging. Some perturbative and non-perturbative effects might have different kinematic dependence and result in intricate effects on the top quark mass in different parts of the phase-space. Using the most precise measurements at $\sqrt{s}=7$ and 8 TeV in CMS (i.e. ideogram in lepton+jets), the top quark mass vs different variables that are sensitive to colour connection, initial/final-state radiation, b-quark kinematics and other effects is studied. One such distribution is displayed in Figure \ref{fig6} which is the opening angle between the jets from the hadronic W in the event. With the current precision, no mismodelling effect has been observed in any of the distributions \cite{mtideo8}.


Measurements with different or independent systematic uncertainties or with different $m_t$ definitions have also been made by CMS. 
Top quark mass is obtained from $t\overline{t}$ cross section \cite{alphasmtpole}, B-hadron lifetime \cite{bhadron}, and kinematic endpoints \cite{endpoint}. 
In addition, $J/\psi$ peak was reconstructed for the first time in $t\overline{t}$ events \cite{jpsi}. Using b-jets from  $J/\psi$ in  $t\overline{t}$ events, the top quark mass can be determined  \cite{kharchilava}  given more precise fragmentation functions and a large number of events. Such a measurement can provide a more realistic assessment of some of the systematic uncertainties in standard top quark mass measurements. 
It is not straightforward to compare these measurements to the direct measurements, however, it is seen that numerically these alternative measurements give compatible values. 

\begin{figure}[h!] 
\centerline{\includegraphics[width=7.5cm]{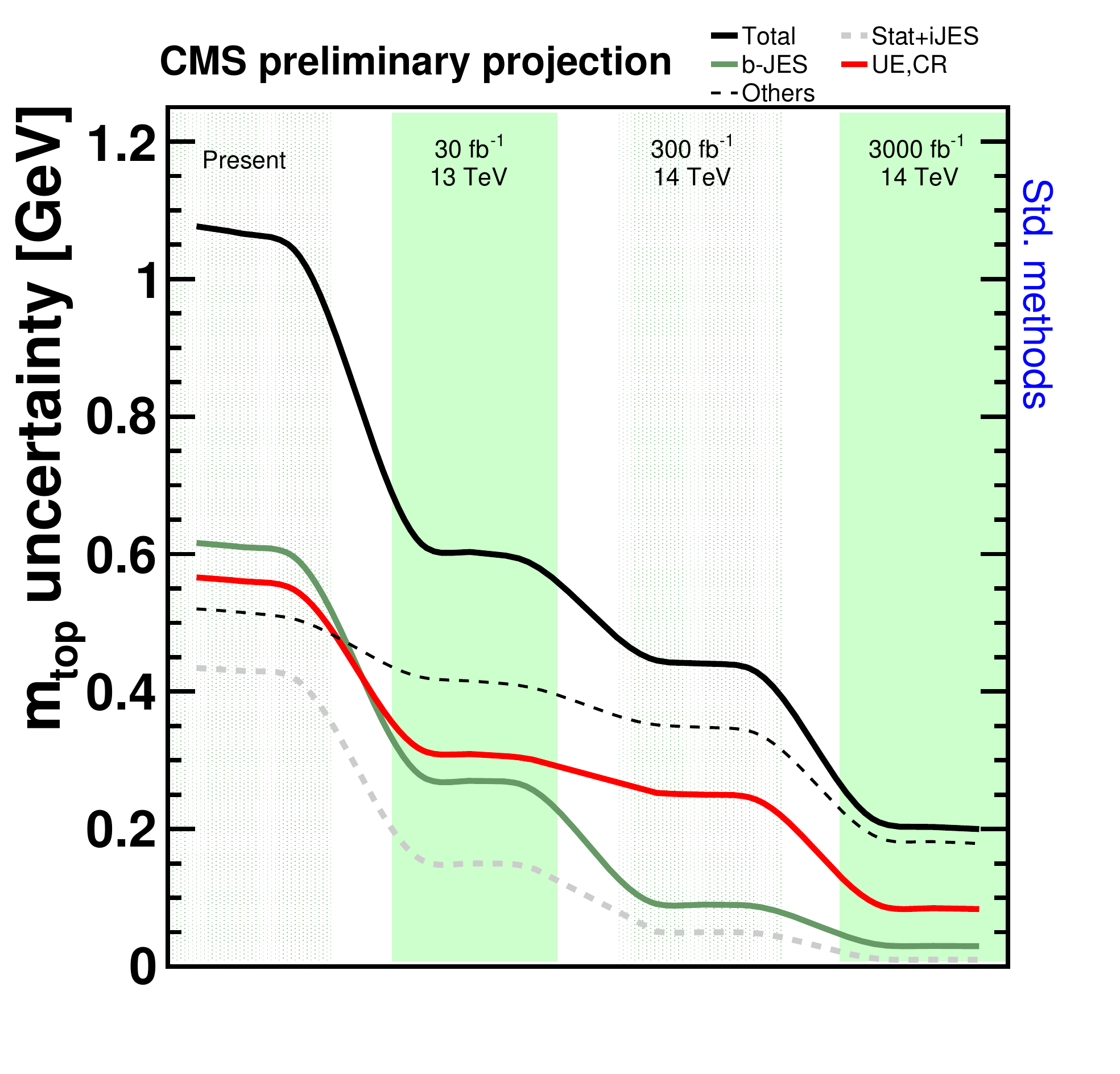}}
\caption{\scriptsize Top-quark-mass precision (in GeV) obtained with standard methods based on reconstruction of the invariant mass of top-quark decay products predicted for various integrated luminosities. The evolution of the total uncertainty as well as selected sources of uncertainties are shown separately. }
\label{fig7} 
\end{figure}

\begin{figure}[h!] 
\centerline{\includegraphics[width=7.5cm]{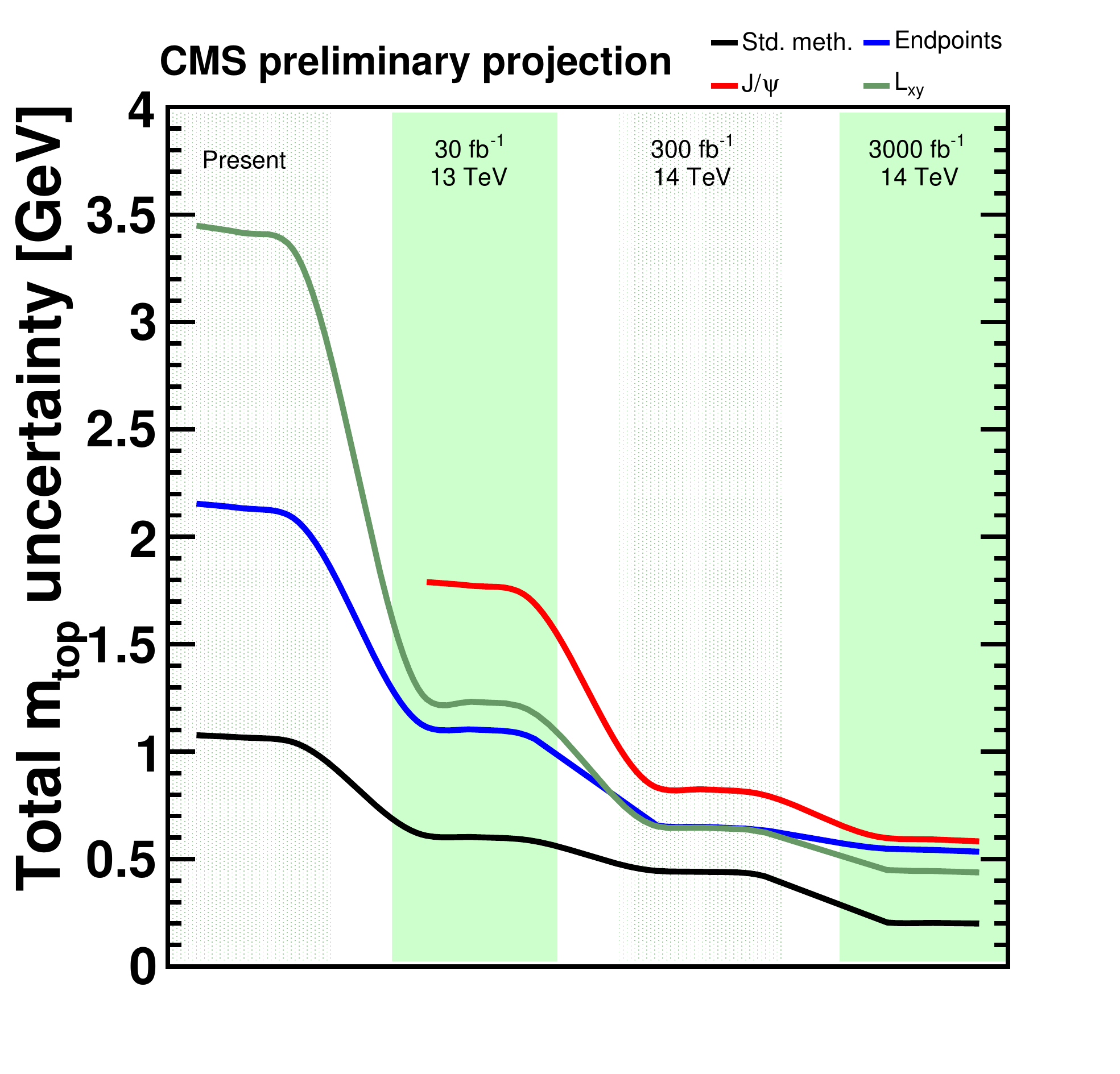}}
\caption{\scriptsize Projection of the top-quark-mass precision (in GeV) obtained with different methods, for various integrated luminosities.}
\label{fig8} 
\end{figure}

Adopting basic optimistic assumptions projections for upgraded LHC up to the integrated luminosities of 3000 $fb^{-1}$ have been made \cite{mtprojection}. With the large datasets of the future LHC runs, improvements in experimental and theoretical systematic uncertainties are expected. Moreover, no limiting irreducible uncertainty is known currently. The total uncertainty at $\sqrt{s}=14$ TeV with 3000 $fb^{-1}$ data, we expect to have a total uncertainty of $\sim0.2$ GeV  for the direct $m_t$ measurements, and $\sim0.6$ GeV for the alternative measurements. The expected evolution of uncertainties in the standard top quark mass measurements is shown in Figure \ref{fig7} and the total uncertainty for the standard methods, along with the alternative methods are shown in Figure \ref{fig8}.

%

\nin
\section{Conclusions}
\nin
Top quark plays an important role in testing and understanding perturbative and non-perturbative QCD. A selected set of top quark measurements made by CMS is presented. A precise determination of the strong coupling constant and top quark pole mass is made using the $t\overline{t}$ cross section and also used for constraints to gluon PDFs at high $x$. Top quark pair differential cross section measurements are used to test SM predictions up to approximate NNLO. Initial/final state modelling is studied with $t\overline{t}$+0 and $>0$ jet events. The top quark mass is measured with a precision better than 1 GeV. Top quark mass related variables are studied as a function of event kinematics to improve our  understanding of our measurements and interpretation. Finally, alternative top quark mass measurements that have different systematic uncertainties or different top quark mass definitions are described. 
\section*{Acknowledgements}
E.Y. would like to thank the organisers for their hospitality. 
\nin














\end{document}